\documentclass[preprint,twoside,web]{ieeecolor}
\usepackage[utf8]{inputenc}

\usepackage{tmi}
\usepackage{cite}
\usepackage{amsmath,amssymb,amsfonts}
\usepackage{graphicx}
\usepackage{textcomp}
\usepackage{subfig}
\usepackage{dsfont}
\usepackage{color, soul}
\usepackage{hyperref}
\usepackage{multirow}
\usepackage{algorithmicx,algpseudocode, algorithm}
\usepackage{booktabs}
\usepackage{float}
\usepackage{graphicx}

\hypersetup{
colorlinks=true,
linkcolor=black
}

\makeatletter
\usepackage{xspace}
\def\@onedot{\ifx\@let@token.\else.\null\fi\xspace}
\DeclareRobustCommand\onedot{\futurelet\@let@token\@onedot}

\newcommand{\algoref}[1]{Alg\onedot~\ref{#1}}


\def\BibTe{{\rm B\kern-.05em{\sc i\kern-.025em b}\kern-.08em T\kern-.1667em\lower.7ex\hbox{E}\kern-.125em x}}

\begin{document}
\title{Guided MRI Reconstruction via Schrödinger Bridge}
\author{Yue Wang, Yuanbiao Yang, Zhuo-xu Cui, Tian Zhou, Bingsheng Huang, Hairong Zheng, \IEEEmembership{Senior Member, IEEE}, Dong Liang, \IEEEmembership{Senior Member, IEEE},   Yanjie Zhu, \IEEEmembership{Senior Member, IEEE}
\thanks{Yue Wang, Yuanbiao Yang, Zhuo-xu Cui contributed equally to this manuscript.}
\thanks{Corresponding author: yj.zhu@siat.ac.cn (Yanjie Zhu).}
\thanks{Yue Wang, Yuanbiao Yang,Tian Zhou, Hairong Zheng, Dong Lian, and Yanjie Zhu are with Lauterbur Research Center for Biomedical Imaging, Shenzhen Institute of Advanced Technology, Chinese Academy of Sciences, Shenzhen, China.}
\thanks{Tian Zhou is with the School of Artificial Intelligence, the University of Chinese Academy of Sciences, Beijing, China.}
\thanks{Zhuo-xu Cui and Dong Liang are with Research Center for Medical AI, Shenzhen Institutes of Advanced Technology, Chinese Academy of Sciences, Shenzhen, China.}
\thanks{Dong Liang is also with Pazhou Lab, Guangzhou, China.}
\thanks{Yue Wang and Bingsheng Huang are with the School of Biomedical Engineering, Shenzhen University Medical School, Shenzhen University,  Shenzhen, China.}
}
\maketitle

\begin{abstract}
Magnetic Resonance Imaging (MRI) is an inherently multi-contrast modality, where cross-contrast priors can be exploited to improve image reconstruction from undersampled data. Recently, diffusion models have shown remarkable performance in MRI reconstruction. However, they still struggle to effectively utilize such priors, mainly because existing methods rely on feature-level fusion in image or latent spaces, which lacks explicit structural correspondence and thus leads to suboptimal performance. To address this issue, we propose $\mathbf{I}^2$SB-Inversion, a multi-contrast guided reconstruction framework based on the Schrödinger Bridge (SB). The proposed method performs pixel-wise translation between paired contrasts, providing explicit structural constraints between the guidance and target images. Furthermore, an Inversion strategy is introduced to correct inter-modality misalignment, which often occurs in guided reconstruction, thereby mitigating artifacts and improving reconstruction accuracy. Experiments on paired T1- and T2-weighted datasets demonstrate that $\mathbf{I}^2$SB-Inversion achieves a high acceleration factor of up to 14.4× and consistently outperforms existing methods in both quantitative and qualitative evaluations.

\end{abstract}

\begin{IEEEkeywords}
Schrödinger Bridge, MRI, Image Reconstruction, Inverse Problem
\end{IEEEkeywords}

\section{Introduction}

\IEEEPARstart{M}{agnetic} Resonance Imaging (MRI) is a powerful imaging technique, but suffers from long acquisition times. A widely adopted strategy to mitigate this limitation is k-space undersampling, followed by image reconstruction that leverages prior information. Typical priors include sparsity and low-rankness of the target images in either image or transform domains, as well as deep priors learned from historical images. Such priors have been extensively explored in compressed sensing (CS) \cite{bilgic2011multi, zhao2012image,cgspirit}  and deep learning (DL) \cite{zhang2018ista,cui2021equilibrated,modl, gungor2023adaptive,cui2023spirit,jiang2024ppn,vn} -based reconstruction methods, leading to significant improvements in reconstruction quality.

Beyond these priors, another important source of prior information can be derived from other MR contrast images of the same subject \cite{gong2015promise}. These images preserve consistent anatomical structures and can help mitigate the degradation caused by undersampling. Approaches exploiting such multi-contrast priors can be broadly classified into two categories: joint reconstruction and guided reconstruction. Joint reconstruction simultaneously reconstructs all multi-contrast images from undersampled k-space data by leveraging their correlations. It is commonly used in imaging techniques inherently designed for multi-contrast acquisitions, such as quantitative MRI \cite{bian2024diffusion} and contrast-enhanced dynamic imaging \cite{van1993keyhole, hess1999maximum, lingala2011accelerated,liang1994efficient,mistretta2006highly}. In these techniques, all contrast images are acquired using a single sequence, usually with the same acceleration factor but different sampling masks, which provides complementary information across images. The priors from multi-contrasts can be jointly modeled for image reconstruction, for example, through joint sparsity \cite{kopanoglu2020simultaneous,trzasko2011group,guo2023joint}, joint low-rankness \cite{haldar2013improved, bustin2019high,bilgic2018improving}, and joint learning-based priors including joint variational networks (JVN) \cite{polak2020joint}, dual-domain reconstruction networks\cite{dual_domain}, and information-sharing networks\cite{sun2019deep}.
However, lacking high-quality references, joint reconstruction methods struggle to achieve high-fidelity reconstruction when the number of available contrasts is limited. As a result, their performance is constrained when applied to conventional single-contrast images, such as T1- and T2-weighted imaging. In contrast, guided reconstruction improves the reconstruction of the target image by introducing a high-quality guidance image as a structural prior. These approaches can be grouped into anatomical prior-based methods \cite{qu2014magnetic,fessler1992regularized,haldar2008anatomically,somayajula2010pet}, deep learning-based methods\cite{yang2020model,Dar2020 }, and multi-modality extensions\cite{knoll2016joint}, where the guidance image is either fully sampled or reconstructed from lightly undersampled data. Owing to the high reliability of the guidance image, it offers a more flexible strategy than joint reconstruction.  

  In recent years, diffusion models have demonstrated remarkable capabilities in MRI reconstruction and achieved state-of-the-art (SOTA) performance \cite{scoremri,peng2022towards}. Despite their advances, only a limited number of studies have explored their application in guided reconstruction due to challenges in effectively integrating priors from the guidance image to the target image. Typically, the guidance prior is incorporated through feature-level fusion in the image or latent space. However, lacking explicit structural correspondence, these approaches impose only weak constraints between the guidance and target images, thereby limiting their performance. To address this limitation, we introduce a nonlinear extension of diffusion models—Schrodinger Bridge (SB) \cite{liu20232, kim2023unpaired,mirza2024super,hu2024adobi} for diffusion-based guided reconstruction. Unlike conventional diffusion models that start from a Gaussian distribution, SB enables general transformations between arbitrary data distributions. Moreover, it allows for pixel-level translation between paired images, which offers an explicit structural constraint between the guidance and target images. Therefore, it may overcome the limitations of feature-level fusion and improve the performance of guided reconstruction.

Another challenge in guided reconstruction is anatomical misalignment across multi-contrast images \cite{LAI201795,9288664,LAI201693}, which arises from subject motion or differences in tissue contrast between imaging modalities. Such misalignment may introduce spurious structures and artifacts in the reconstructed image. To address this, we incorporate an Inversion strategy inspired by image editing techniques \cite{huberman2024edit}, which reverses the generative process to recover latent variables from a known image. In our SB-based guided reconstruction, Inversion is used to identify the guiding variable along the SB trajectory that best corresponds to the target image, correcting reconstruction errors caused by misalignment between multi-contrast images.

In this work, we propose a novel SB-based guided reconstruction framework with Inversion to correct spatial misalignment between multi-contrast images. First, an initial target image is generated by translating the guidance image through SB sampling while enforcing data consistency with the acquired k-space. This image then serves as the starting point for Inversion to infer a structurally aligned guidance variable, which is subsequently used for a second round of SB sampling to obtain the final reconstruction. The main contributions are:

1) We propose a Schrodinger Bridge (SB)-based framework for guided MRI reconstruction, enabling pixel level cross-modality mapping. 

2) We introduce an image-domain Inversion strategy along the SB path, which searches for approximately aligned guidance variables based on the initial reconstruction result to mitigate the effects of spatial misalignment. 

3) We demonstrate that the proposed $\mathrm{I}^2$SB-Inversion achieves superior reconstruction quality compared with SOTA methods under high acceleration.

The following sections of the paper are organized as follows: Section \ref{Background} introduces the background, Section \ref{theory and methods} describes the proposed method, and Section \ref{experiments} provides the experimental results. Discussion and conclusion are given in Section \ref{discussion} and Section \ref{conclusions}.

\section{Background}\label{Background}

\subsection{Score-based Generative Model (SGM)}
SGM\cite{ho2020denoising,song2019generative,song2020score} is a general framework for diffusion models that describes the diffusion process as the solution of stochastic differential equations (SDEs). Given an initial state \( \mathbf{x}_0 \) sampled from a distribution \( p_0 \), SGM constructs a forward diffusion process \( \{ \mathbf{x}_t \}_{t=0}^1 \), which gradually transfers \( \mathbf{x}_0 \) into a Gaussian distribution via the forward SDE\cite{song2020score} as:

\begin{equation}
    \mathrm{d} \mathbf{x}_t=f_t \mathrm{d} t+g_t \mathrm{d} \mathbf{w}_t,
    \label{SDE}
\end{equation}

\noindent where \( f_t \) is the drift function of \( \mathbf{x}_t\) and \( g_t \) is a scalar function known as the diffusion coefficient. 
\( \mathbf{w}_t \) is the standard Wiener process. A concrete example is the Denoising Diffusion Probabilistic Model (DDPM)\cite{ho2020denoising}, which can be viewed as a discrete-time approximation of the Variance Preserving SDE (VP-SDE), the continuous formulation of this process, with:
\begin{equation}
f_t = -\frac{1}{2} \beta(t) \mathbf{x}_t, \quad g_t = \sqrt{\beta(t)},
\end{equation}
where $\beta(t)$ is a monotonically increasing noise schedule. This design ensures a smooth transition from the data distribution to a Gaussian distribution.

In contrast, the forward process can be reversed by the following reverse-time SDE\cite{song2020score} from a Gaussian distribution to a target distribution:
\begin{equation}
    \mathrm{d} \mathbf{x}_t=\left[f_t-g_t^{2} \nabla_{\mathbf{x}} \log p_{t}(\mathbf{x})\right] \mathrm{d} t+g_t \mathrm{d} \mathbf{\bar w}_t,
    \label{related-work reverse sde}
\end{equation}

\noindent where \( \nabla_{\mathbf{x}} \log p_t(\mathbf{x}) \) denotes the score function of \( p_t \), and \( \mathbf{\bar w}_t \) is the standard Wiener process when time goes backward from $t=1$ to $t=0$\cite{song2019generative}. Then starting from Gaussian noise, we can obtain samples \( \mathbf{x}_0 \sim p_0 \) through Eq.~\eqref{related-work reverse sde}.

\subsection{Schrödinger Bridge}
SB is a nonlinear extension of score-based diffusion models that enables transformations between two arbitrary data distributions~\cite{schrodinger1932theorie,leonard2013survey,chen2021likelihood,fernandes2022shooting}. Unlike conventional diffusion models\cite{ho2020denoising,song2019generative}  that are initialized from Gaussian noise with no structural information, SB can be initialized from related images that retain meaningful structural features. This relaxation of the Gaussian prior assumption allows SB to better exploit available prior information and is particularly suited for conditional generation~\cite{chen2021likelihood,fernandes2022shooting}. A recent work in MRI reconstruction, i.e., the Fourier-constrained diffusion bridge (FDB)~\cite{fdb}, adapts SB to directly learn the mapping between undersampled and fully sampled data and demonstrates its benefit. In guided reconstruction, SB offers a direct and convenient way to translate between guidance and target images by learning the mapping between their distributions. However, the original SB model is computationally complex and difficult to apply in practice. Therefore, we employ a more practical variant developed for image-related tasks, namely image-to-image SB ($\mathrm{I^2SB}$)\cite{liu20232} in our guided reconstruction framework.

$\mathrm{I^2SB}$ adopts a similar procedure of the DDPM\cite{ho2020denoising} for training and generation. Specifically, given $\mathbf{x}_0 \sim p_{\text{tar}}$ and $\mathbf{x}_N \sim p_{\text{guid}}$, where $p_{\text{tar}}$ and $p_{\text{guid}}$ denote the distributions of the target and guidance images, respectively, the forward process can be formed as:

\begin{equation}
q(\mathbf{x}_n|\mathbf{x}_0,\mathbf{x}_N) = \mathcal{N}(\mathbf{x}_n;\,\mu_n(\mathbf{x}_0,\mathbf{x}_N),\Sigma_n),
\label{forward_pro}
\end{equation}
with
\begin{equation}
\mu_n = \frac{\bar{\sigma}_n^2}{\bar{\sigma}_n^2 + \sigma_n^2}\mathbf{x}_0 
      + \frac{\sigma_n^2}{\bar{\sigma}_n^2 + \sigma_n^2}\mathbf{x}_N, 
\quad 
\Sigma_n = \frac{\eta^2 \sigma_n^2\bar{\sigma}_n^2}{\bar{\sigma}_n^2 + \sigma_n^2}I,
\end{equation}
where $\sigma_n^2 = \sum_{i=0}^{n} \beta_i$ and $\bar{\sigma}_n^2 = \sum_{i=n+1}^{N} \beta_i$ denote the accumulated noise variances from each side, $\beta_i$ represents the noise schedule, and $\eta \geq 0$ is a variance scaling parameter. In the training phase, given a training image pair $(\mathbf{x}_0, \mathbf{x}_N)$,  the intermediate state $\mathbf{x}_n$ at time step $n$ is sampled according to Eq.~\eqref{forward_pro}. Then the residual predictor $\epsilon_\theta(\mathbf{x}_n, n; \theta)$ is trained to predict the residual noise in $\mathbf{x}_n$. The training algorithm is illustrated in Alg.~\ref{alg:training}.


In the generation phase, the posterior sampling between adjacent steps can be derived using the Chapman-Kolmogorov~\cite{schrodinger1932theorie,leonard2013survey} relation as:
\begin{align}
q(\mathbf{x}_n \mid \mathbf{x}_0, \mathbf{x}_N)
= \int p(\mathbf{x}_n \mid \mathbf{x}_0, \mathbf{x}_{n+1})\,
q(\mathbf{x}_{n+1} \mid \mathbf{x}_0, \mathbf{x}_N)\, d\mathbf{x}_{n+1}.
\label{eq:ck_relation}
\end{align}

Since both $p$ and $q$ follow Gaussian distributions, the above integration also yields a Gaussian form.  
By matching the mean and variance terms, the closed-form expression of the posterior sampling can be obtained as:  
\begin{align}
    p(\mathbf{x}_{n} \mid\mathbf{x}_0 , \mathbf{x}_{n+1}) 
    &= \mathcal{N}\left(\mathbf{x}_{n}; \frac{\alpha_n^2}{\alpha_n^2 + \sigma_n^2} \mathbf{x}_0  + \frac{\sigma_n^2}{\alpha_n^2 + \sigma_n^2} \mathbf{x}_{n+1}, \right. \nonumber \\
    &\quad \left. \frac{ \eta^2   \sigma_n^2 \alpha_n^2}{\alpha_n^2 + \sigma_n^2} \cdot I\right),
    \label{posterior_clean}
\end{align}
where $\alpha_n^2 = \sigma_{n+1}^2-\sigma_{n}^2 = \beta_n$. As $\mathbf{x}_0$ is unavailable, $\epsilon_\theta(\mathbf{x}_n, n; \theta)$ is used to predict the residual noise and derive an approximation  of the target image denoted as $\mathbf{x}_0^\epsilon$, which is then used for posterior sampling:



\begin{align}
    p(\mathbf{x}_{n} \mid\mathbf{x}_0^\epsilon, \mathbf{x}_{n+1}) 
    &= \mathcal{N}\left(\mathbf{x}_{n}; \frac{\alpha_n^2}{\alpha_n^2 + \sigma_n^2} \mathbf{x}_0^\epsilon + \frac{\sigma_n^2}{\alpha_n^2 + \sigma_n^2} \mathbf{x}_{n+1}, \right. \nonumber \\
    &\quad \left. \frac{ \eta^2   \sigma_n^2 \alpha_n^2}{\alpha_n^2 + \sigma_n^2} \cdot I\right).
    \label{posterior}
\end{align}


Based on the above derivation, the generation process begins with $\mathbf{x}_N$ and iteratively performs residual prediction, image generation, and posterior sampling, allowing progressive refinement toward high-quality image translation.

\begin{algorithm}[H]
\caption{\textbf{Training}}
\label{alg:training}
\textbf{Input:}\\
$p_\text{tar}$: target distribution\\
$p_\text{guid}$: guided distribution\\
$\epsilon_\theta(\cdot, \cdot; \theta)$: residual predictor\\[0.3em]
\textbf{Output:}\\
Trained parameters $\theta$
\begin{algorithmic}[1]
\Repeat
    \State Sample $n \sim \mathcal{U}([0, N])$, $\mathbf{x}_0 \sim p_\text{tar}(\mathbf{x}_0)$, $\mathbf{x}_N \sim p_\text{guid}(\mathbf{x}_N|\mathbf{x}_0)$
    \State Sample $\mathbf{x}_n \sim q(\mathbf{x}_n | \mathbf{x}_0, \mathbf{x}_N)$ 
    \State Take a gradient descent step on
    \[
        \left\lVert \epsilon_\theta(\mathbf{x}_n, n; \theta) - \frac{\mathbf{x}_n - \mathbf{x}_0}{\sigma_n} \right\rVert
    \]
\Until{convergence}
\end{algorithmic}
\end{algorithm}





\section{Theory and methods}\label{theory and methods}
\begin{figure*}[!t]
    \centerline{\includegraphics[width=\textwidth]{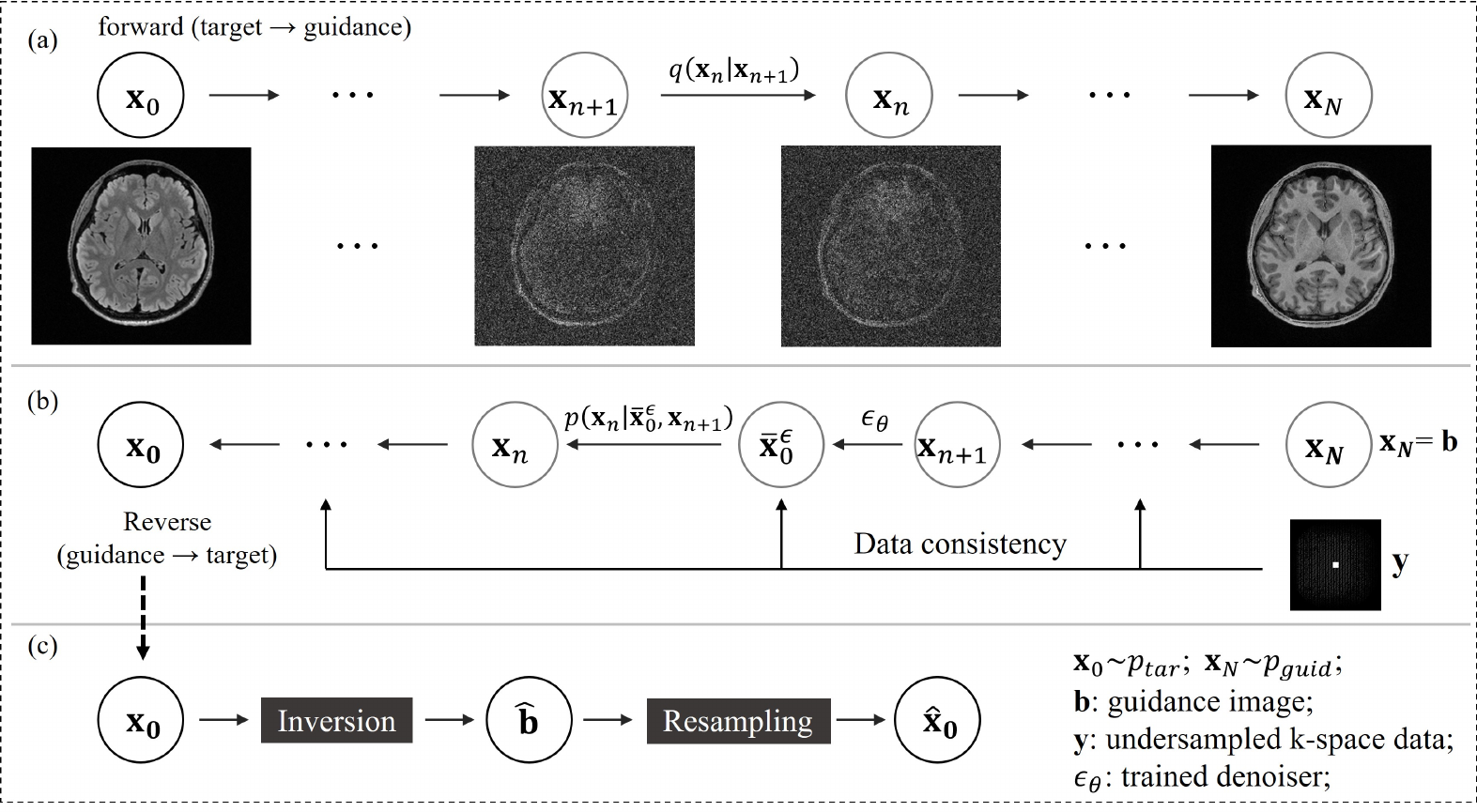}}
    \caption{(a) Forward step: The target image $\mathbf{x}_0$ gradually transforms into the guidance image $\mathbf{x}_N$, forming the SB trajectory $\{\mathbf{x}_0, \dots, \mathbf{x}_N\}$. Paired data $(\mathbf{x}_n$, $\mathbf{x}_0)$ are then used to train the denoiser $\epsilon_\theta$.
(b) Backward step: Starting from $\mathbf{x}_N$, the target image $\mathbf{x}_0$ is reconstructed through iterative sampling, with data consistency enforced at each step to match the acquired k-space data.
(c) Inversion step: After (b), $\mathbf{x}_0$ is converted into the aligned guidance image $\hat{\mathbf{b}}$, followed by resampling to obtain the final reconstruction $\hat{\mathbf{x}}_0$.}
    \label{fig:general}
\end{figure*}

\subsection{Guided MRI Reconstruction}
The imaging model of MR reconstruction can be formulated as:
\begin{equation}
    \mathbf{y}=\mathbf{A}\mathbf{x}+\boldsymbol{\xi},
    \label{MR forward model}
\end{equation}
where $\mathbf{y}$ is the undersampled k-space data, $\mathbf{x}$ is the image to be reconstructed, $\mathbf{A}$ denotes the encoding matrix, $\mathbf{A}=\mathbf{M}\cdot \mathbf{F} \cdot \text{csm}$, $\mathbf{M}$ is the undersampling operator, $\mathbf{F}$ denotes the Fourier operator, $\text{csm}$ denotes the coil sensitivity, and ${\boldsymbol{\xi}} \sim \mathcal{N}(0, \sigma^2_{\boldsymbol{\xi}})$. For 2D image, $\mathbf{x} \in \mathbb{C}^n$, $\mathbf{y} \in \mathbb{C}^m$ and $\mathbf{A} \in \mathbb{C}^{m\times n}$.

Since Eq.~\eqref{MR forward model} is an ill-posed problem, additional prior information is necessary to solve it. Given a guidance image \(\mathbf{b}\), incorporating the structural similarity prior between the guidance image and the reconstructed image, the solution to Eq.~\eqref{MR forward model} can be expressed as the following constrained optimization problem:


\begin{equation}
\min_{\mathbf{x}} R(\mathbf{x},\mathbf{b})
\quad \text{subject to} \quad \mathbf{y} = \mathbf{A}\mathbf{x},
\label{MR optimization problem}
\end{equation}

\noindent where $\mathbf{b}$ is the guidance image, and $R(\mathbf{x}, \mathbf{b})$ represents the structural similarity prior between $\mathbf{x}$ and $\mathbf{b}$. In traditional methods, the optimization problem is often decoupled into two alternately optimized subproblems to obtain the optimal solution.



\begin{algorithm}[H]
\caption{\textbf{I$^2$SB-Recon}}
\label{alg:ddpm_guide}
\textbf{Input:} 
\\$\mathbf{b} \sim p_{\text{guid}}$: \text{guidance image}
\\ $\epsilon_\theta(\cdot, \cdot; \theta)$: residual predictor\\
$\mathbf{y}$: undersampled k-space data \\
\textbf{Output:} 
\\$\mathbf{x}_0$: Reconstructed image 
\begin{algorithmic}[1]
\\$\mathbf{x}_N=\mathbf{b}$
\For{$n = N-1,\dots,1,0$}
    \State $\mathbf{x}_0^{\epsilon} = \mathbf{x}_n - \epsilon_\theta(\mathbf{x}_n, n; \theta)\, \sigma_n$ \Comment{estimate target image}
    \State $\mathbf{\bar{x}}_0^\epsilon=\text{CG}(\mathbf{x}_0^\epsilon,\mathbf{y})$ \Comment{enforce k-space data consistency}
    \State Sample $\mathbf{x}_{n} \sim p(\mathbf{x}_{n} \mid \mathbf{\bar{x}}_0^\epsilon, \mathbf{x}_{n+1})$
\EndFor
\State \textbf{return} $\mathbf{x}_0$
\end{algorithmic}
\end{algorithm}

\subsection{Guided Reconstruction Based on Schrödinger Bridge}


The regularizer $R(\mathbf{x}, \mathbf{b})$ can be interpreted as the evolution path between the distributions of target and guidance images, denoted as $p_{\text{tar}}$ and $p_{\text{guid}}$, respectively. Within the SB framework, this path can be formalized as a probabilistic process $\{\mathbf{x}_n\}_{n=0}^N$ that evolves from $p_{\text{tar}}$ to $p_{\text{guid}}$. In the reverse process, following the framework of $\mathrm{I^2SB}$, a network $\epsilon_\theta$ is used to predict the residual noise $\|\mathbf{x}_n - \mathbf{x}_0\|$ from $\mathbf{x}_n$. This network $\epsilon_\theta$ is trained according to Alg.~\ref{alg:training} using paired guidance and target MR image datasets. 

After training, the target image can be reconstructed through iterative sampling $\{\mathbf{x}_n\}$, $n = N \to 0$. Starting from $n = N$ and $\mathbf{x}_N = \mathbf{b}$, the initial estimation of $\mathbf{x}_0$ can be represented by

\begin{equation}
\mathbf{x}_0^{\epsilon} = \mathbf{x}_n - \epsilon_\theta(\mathbf{x}_n, n; \theta)\, \sigma_n.
\label{eq:x0_estimation}
\end{equation}

Then, the data consistency constraint in Eq.~\eqref{MR optimization problem} should be applied to ensure that the generated image is consistent with the acquired k-space data. Accordingly, Eq.~\eqref{MR optimization problem} can be reformulated as an optimization problem,

\begin{equation}
\min_{\mathbf{x}} \|\mathbf{y}-\mathbf{A}\mathbf{x}\|_2^2 + \lambda \|\mathbf{x} - \mathbf{x}^\epsilon_0\|_2^2,
\label{MR reverse problem}
\end{equation}
where $\lambda$ is the regularization parameter. Various methods can be used to solve this optimization problem, including the projection method~\cite{griswold2002generalized}, gradient descent method~\cite{lustig2007sparse}, and conjugate gradient (CG)~\cite{pruessmann1999sense}. Previous studies~\cite{chung2023decomposed,zhu2023denoising} have demonstrated that, under the assumption that the tangent space of the denoised sample $\mathbf{x}_0^{\epsilon}$ can be represented as a Krylov subspace, using the standard CG method ensures that the optimization direction aligns with the gradient direction of the distribution. This property keeps the reconstructed image well confined within the desired distribution $p_{\text{tar}}$. Therefore, we employ the CG method to solve the optimization problem in this study. Its solution is denoted as 
\begin{align}
\mathbf{\bar{x}}_0^\epsilon=\mathrm{CG}(\mathbf{x}_0^{\epsilon}, \mathbf{y}).
\label{correct}
\end{align}

Since the CG correction remains on $p_{\text{tar}}$ and satisfies the SB prior, iterations along the SB path can be performed to progressively refine the reconstructed image. We employ posterior sampling to map the CG-corrected sample back onto the SB path for the next iteration:


\begin{align}
p(\mathbf{x}_{n} \mid \mathbf{\bar{x}}_0^\epsilon, \mathbf{x}_{n+1})
&= \mathcal{N}\left(\mathbf{x}_{n}; \frac{\alpha_n^2}{\alpha_n^2 + \sigma_n^2} , \mathbf{\bar{x}}_0^\epsilon + \frac{\sigma_n^2}{\alpha_n^2 + \sigma_n^2} \mathbf{x}_{n+1}, \right. \nonumber \
\\&\quad \left. \frac{\eta^2 \sigma_n^2 \alpha_n^2}{\alpha_n^2 + \sigma_n^2} \cdot I \right).
\label{correct}
\end{align}


 \algoref{alg:ddpm_guide} illustrates the reconstruction algorithm, termed $\mathrm{I}^2$SB-Recon.

\subsection{Guided Reconstruction under the Inversion}

In guided reconstruction, the guidance and target images are typically assumed to be perfectly aligned in the spatial domain. However, this assumption often fails in real-world scenarios due to subject motion during scanning, leading to misalignment between multi-contrast images and degradation in the reconstructed image quality. To mitigate this issue, an Inversion strategy is introduced. Specifically, using the reconstructed image $\mathbf{x}_0$ obtained by I$^2$SB-Recon as an initialization, the Inverse process samples along the SB trajectory toward the guidance distribution to infer a corresponding sample $\hat{\mathbf{b}}$, followed by re-reconstructing the image with $\hat{\mathbf{b}}$ as the new guidance. Since $\mathbf{x}_0$ has been partially corrected by the acquired data, $\hat{\mathbf{b}}$ is expected to be more spatially aligned with the true target image than the original guidance image $\mathbf{b}$, thereby alleviating reconstruction errors caused by misalignment.


In the Inversion process, a deterministic Probability Flow ODE is adopted instead of the SDE to avoid introducing additional randomness. 
By setting $\eta = 0$ in Eq.~\eqref{posterior_clean}, the posterior sampling process degenerates into an ODE, simplifying the sampling procedure as follows:

\begin{equation}
    \mathbf{x}_{n} = \frac{\alpha_n^2}{\alpha_n^2 + \sigma_n^2}  \mathbf{x}_0 + \frac{\sigma_n^2}{\alpha_n^2 + \sigma_n^2}  \mathbf{x}_{n+1}, 
    \label{flow}
\end{equation}

\noindent $\mathbf{x}_0$ in the above formula can be approximated using Eq.~\eqref{eq:x0_estimation} at step $n\!+\!1$, i.e. $\mathbf{x}_0^{\epsilon} = \mathbf{x}_{n+1} - \epsilon_\theta(\mathbf{x}_{n+1}, n+1; \theta)\, \sigma_{n+1}$, and Eq.~\eqref{flow} can then be reformulated as:




\begin{equation}
    \mathbf{x}_{n+1}
    = \mathbf{x}_n + \frac{\alpha_n^2 \sigma_{n+1}}{\alpha_n^2 + \sigma_n^2}
    \, \epsilon_\theta(\mathbf{x}_{n+1}, n+1; \theta).
    \label{flow1}
\end{equation}

However, during the opposite flow (Inversion process), the residual part $\epsilon_\theta(\mathbf{x}_{n+1}, n+1; \theta)$ is unknown at time step \( n \) when inference from $n$ to $n+1$. A feasible solution is to use $\epsilon_\theta(\mathbf{x}_{n}, n; \theta)$ as a substitute. To achieve this, we reconsider the marginal distribution $q(\mathbf{x}_n \mid \mathbf{x}_0, \mathbf{x}_N)$, which can also be simplified as:
\begin{align}
\mathbf{x}_{n}
&= \frac{\bar{\sigma}_n^2}{\bar{\sigma}_n^2 + \sigma_n^2}\mathbf{x}_0 
   + \frac{\sigma_n^2}{\bar{\sigma}_n^2 + \sigma_n^2}\mathbf{x}_N,
   \label{flow_marginal}
\end{align}

\noindent $\mathbf{x}_n = \mathbf{x}_0^{\epsilon} - \epsilon_\theta(\mathbf{x}_n, n; \theta)\, \sigma_n$ can be derived from  Eq.~\eqref{eq:x0_estimation}. Then it can be combined with Eq.~\eqref{flow_marginal} to obtain the following formula:  


\begin{equation}
\epsilon_\theta(\mathbf{x}_{n}, n; \theta) = -\frac{\sigma_n}{\bar{\sigma}_n^2 + \sigma_n^2} \mathbf{x}_0^\epsilon + \frac{\sigma_n}{\bar{\sigma}_n^2 + \sigma_n^2} \mathbf{x}_N.
\label{sample}
\end{equation}

Connecting Eq.~\eqref{sample} with different time steps $n+1$ and $n$, the relationship between the predicted noises can be expressed as:

\begin{equation}
    \epsilon_\theta(\mathbf{x}_{n+1}, n+1; \theta) = \frac{\sigma_{n+1} }{\sigma_n} \cdot \epsilon_\theta(\mathbf{x}_n, n; \theta).
    \label{relation between noise}
\end{equation}

By substituting Eq.~\eqref{relation between noise} into Eq.~\eqref{flow1}, a feasible formulation for the Inversion strategy can be derived as:

\begin{equation}
   \mathbf{x}_{n+1} = \mathbf{x}_n + \frac{\alpha_n^2 \sigma_{n+1}^2}{\sigma_n (\alpha_n^2 + \sigma_n^2)} \cdot \epsilon_\theta(\mathbf{x}_n, n; \theta),
   \label{eq:inversion}
\end{equation}

\noindent Through Eq.~\eqref{eq:inversion}, the Inversion process can be formalized using $\{\mathbf{x}'_n\}$, $n = 0 \rightarrow N$, 
where $\mathbf{x}'_0$ is reconstructed using I$^2$SB-Recon. 
The pseudocode of I$^2$SB-Inversion is presented in ~\algoref{alg:highSNR}.

\begin{algorithm}[H]
\caption{I$^2$SB-Inversion }
\label{alg:highSNR}
\textbf{Input:} 
\\ $\mathbf{b} \sim p_{\text{guid}}$: guidance image
\\ $\epsilon_\theta(\cdot, \cdot; \theta)$: residual predictor
\\ $\mathbf{y}$: undersampled k-space data 
\\
\textbf{Output:} 
\\ $\hat{\mathbf{x}}_0$: reconstructed image 
\begin{algorithmic}[1]
\State Run Alg.~\ref{alg:ddpm_guide} to obtain $\mathbf{x}'_0= \mathbf{x}_0$
\State \textbf{Inversion process:}
\For{$n = 0, 1, \dots, N-1$}
    \State $\mathbf{x}'_{n+1} = \text{Inversion}(\mathbf{x}'_n)$ using Eq.~\eqref{eq:inversion}
\EndFor
\State $\hat{\mathbf{b}} = \mathbf{x}'_N$

\State Run Alg.~\ref{alg:ddpm_guide} starting from $\hat{\mathbf{b}}$ to obtain final $\hat{\mathbf{x}}_0$
\State \textbf{return} $\hat{\mathbf{x}}_0$
\end{algorithmic}
\end{algorithm}

\begin{figure*}[!t]
    \centerline{\includegraphics[width=1\textwidth]{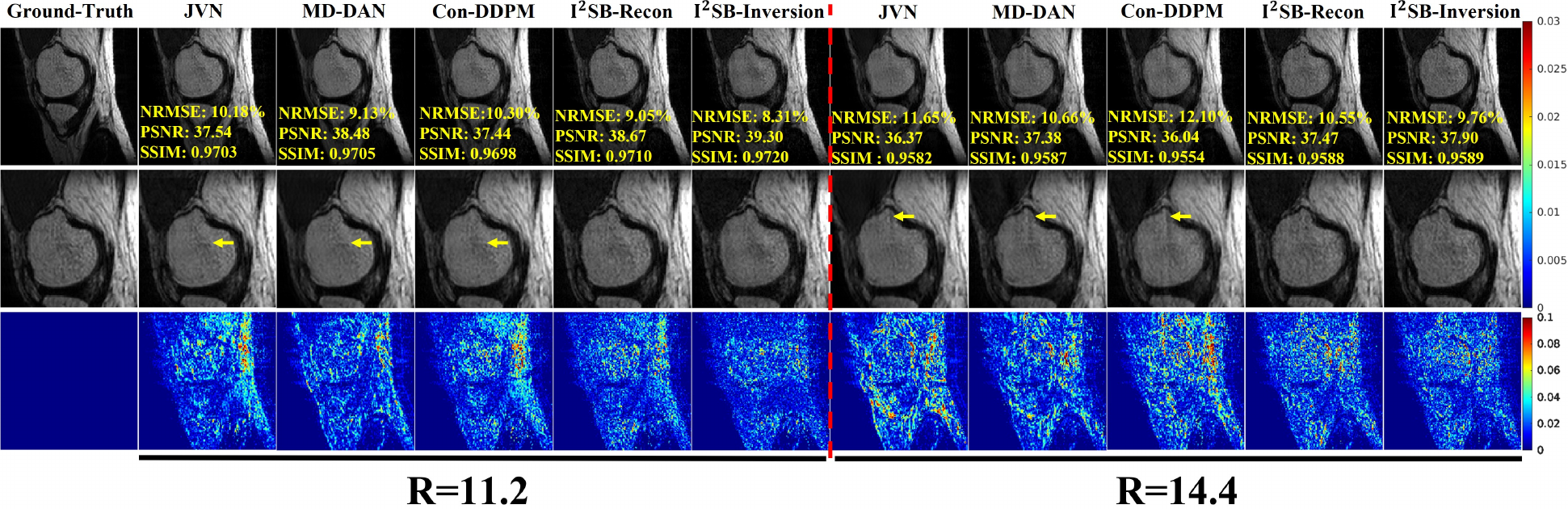}}
    \caption{Guided Reconstruction results in knee dataset at R = 11.2 and R = 14.4 . The top row shows the ground truth and the reconstructions obtained using different methods. The second row shows an enlarged view of the ROI, and the third row displays the error map of the reconstructions. Regions with blurring or artifacts in the compared methods are marked with yellow arrows. }
    \label{fig:Guided_knee}
\end{figure*}

\begin{figure*}[!t]
    \centerline{\includegraphics[width=1\textwidth]{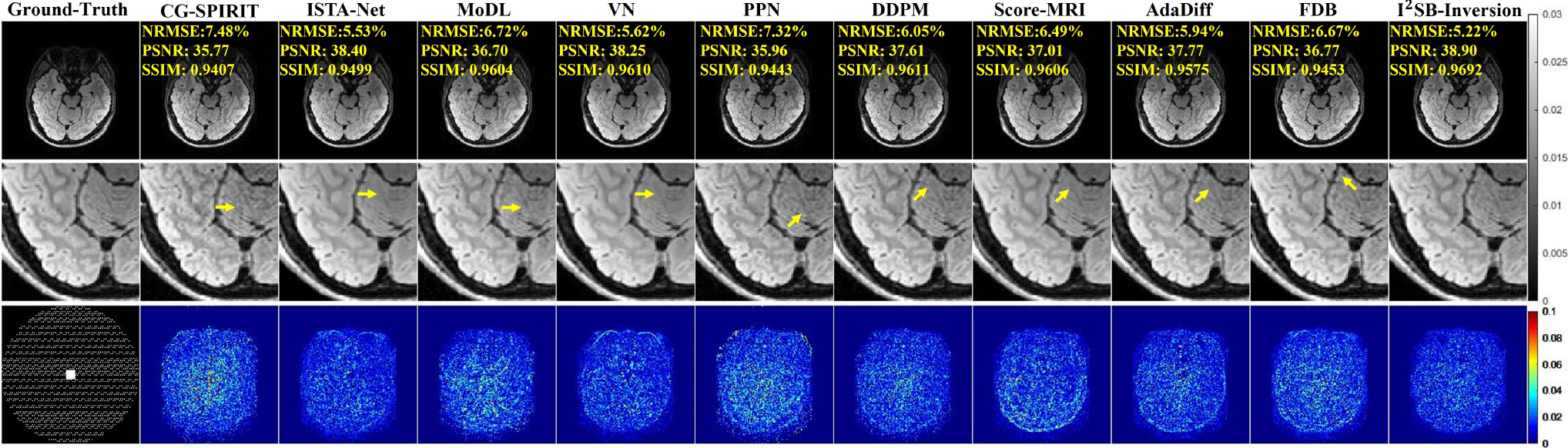}}
    \caption{Comparison between direct reconstruction methods and the proposed $\mathrm{I}^2$SB-Inversion (with guidance) on the brain dataset at R = 11.2. The top row shows the ground truth and the reconstructions obtained using different methods. The second row shows an enlarged view of the ROI, and the third row displays the error map of the reconstructions. Regions with blurring or artifacts in the compared methods are marked with yellow arrows. }
    \label{fig:direct_11.2}
\end{figure*}

\begin{figure*}[!t]
    \centerline{\includegraphics[width=1\textwidth]{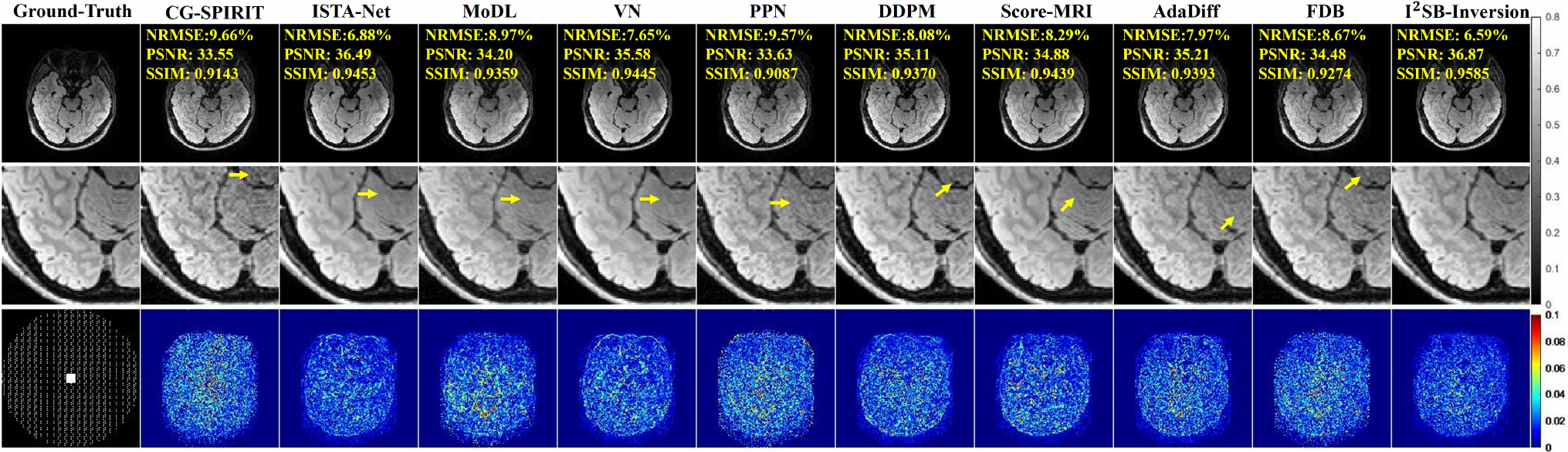}}
    \caption{Comparison between direct reconstruction methods and the proposed $\mathrm{I}^2$SB-Inversion (with guidance) on the brain dataset at R = 14.4. The top row shows the ground truth and the reconstructions obtained using different methods. The second row shows an enlarged view of the ROI, and the third row displays the error map of the reconstructions. Regions with blurring or artifacts in the compared methods are marked with yellow arrows.}
    \label{fig:direct_14.4}
\end{figure*}

\section{Experiments}\label{experiments}
\subsection{Experimental Setup}
\subsubsection{Experimental Data}
We conducted experiments on two datasets comprising paired T1- and T2-weighted images acquired on a 3T MR scanner (uMR 790, United Imaging Healthcare, China): a knee dataset and a brain dataset. All experiments were approved by the local institutional review board. The brain dataset includes fully sampled k-space data from 29 healthy volunteers using a 32-channel head coil. T1-weighted images were acquired with a 3D GRE sequence, and corresponding T2-FLAIR images with a 3D FSE sequence. For each volunteer, both sequences had identical positioning and spatial resolution, with acquisition matrix = 240 $\times$ 240 $\times$ 176 and FOV = 240 $\times$ 240 $\times$ 176 mm$^3$. Imaging parameters were as follows: For the T1 sequence: TR/TE = 7.7/3 ms, flip angle = 9°, echo train length = 176, bandwidth = 250 Hz/pixel, scan time = 8 min 36 s; For the T2 sequence: TR/TE = 6000/396.4 ms, echo train length = 240, bandwidth = 600 Hz/pixel, elliptical scanning, scan time = 14 min. 

The knee dataset includes fully sampled k-space data from 13 healthy volunteers using a 12-channel knee coil. Both T1- and T2-weighted images were acquired using the 3D FSE sequence with acquisition matrix = $240 \times 240 \times 140$ and FOV = $160 \times 160 \times 150~\mathrm{mm}^3$. Imaging parameters were as follows: For the T1 sequence, TR/TE = $600/15.36~\mathrm{ms}$, echo train length = $50$, bandwidth = $450~\mathrm{Hz/pixel}$, scan time = $4~\mathrm{min}~31~\mathrm{s}$; for the T2 sequence, TR/TE = $1400/259.2~\mathrm{ms}$, echo train length = $120$, bandwidth = $400~\mathrm{Hz/pixel}$, scan time = $9~\mathrm{min}~46~\mathrm{s}$.

The 3D k-space data were divided into 2D slices along the RO direction by applying the inverse Fourier transform. For both the brain and knee datasets, background-only slices were discarded—specifically, the first 25 and last 50 slices for the brain data, and the first and last 48 slices for the knee data. Then, coil compression was used to compress the data to 18 channels to reduce computational load \cite{zhang2013coil}. Zero-padding was applied to increase the image size to 256 $\times$ 256, facilitating network operations. In our experiments, the T1-weighted image served as guidance for reconstructing the corresponding T2-weighted image. During training, both T1- and T2-weighted images were fully sampled, whereas during testing, the guidance T1-weighted images remained fully sampled, and the T2-weighted images were retrospectively undersampled with net acceleration factors (R) of 11.2 and 14.4. Note that undersampling masks were defined on the zero-padded 256 $\times$ 256 k-space, and R was computed with respect to this size. For the brain dataset, 23 volunteers were randomly selected for the training set, yielding 4163 matched image pairs, while the remaining 6 volunteers formed the test set, comprising 1086 matched image pairs. For the knee dataset, 10 volunteers were randomly selected for the training set, yielding 1600 matched image pairs, while the remaining 3 volunteers formed the test set, comprising 480 matched image pairs.

The CAIPI undersampling scheme \cite{breuer2005controlled} was employed with a 48$\times$48 k-space center fully sampled. The coil sensitivity maps were estimated using the fully sampled k-space center with the ESPIRiT algorithm \cite{uecker2014eSPIRiT}.



\subsubsection{Implementation Details}

The network $\epsilon(\mathbf{x}_n, n; \theta)$ was 
implemented using a U-Net architecture in\cite{dhariwal2021diffusion}, which is widely used in diffusion models. The network has 552 million parameters across 91 layers and is initialized using the publicly released ADM checkpoint trained on ImageNet 256×256 (\href{https://github.com/openai/guided-diffusion}{official ADM checkpoint}). We refer readers to the original work for detailed configurations. Because MR data are complex-valued, the real and imaginary components were represented as two separate channels in the network. During the data consistency step, they were recombined into a complex form to perform Fourier transforms, thereby enforcing consistency with the acquired k-space data. Implementation details and source code for $\mathrm{I}^2$SB-Inversion are available at: \url{https://github.com/zhyjSIAT/I2SB-Inversion}.

We compared the $\mathrm{I}^2$SB-Inversion method with several SOTA approaches. For guided reconstruction methods, we selected the unrolling-based method MD-DAN \cite{yang2020model}, the conditional distribution learning method Con-DDPM (defined as a conditional DDPM guided by an image), the JVN \cite{polak2020joint} (defined as a conditional VN guided by an image), as well as I$^2$SB-Recon (without Inversion) to compare their performance under different guidance paradigms. 
We also compared our method with four categories of direct reconstruction methods without guidance: (a) the traditional parallel imaging and compressed sensing method CG-SPIRIT \cite{cgspirit}; (b) unrolling-based methods, including ISTA-Net \cite{zhang2018ista} and MoDL \cite{modl}; (c) diffusion-based methods, including DDPM \cite{ho2020denoising}, ScoreMRI \cite{scoremri}, AdaDiff \cite{gungor2023adaptive}, and PPN \cite{jiang2024ppn}; and (d) the Schrödinger bridge–based method FDB \cite{fdb}. The implementation details were as follows. 

For unrolling-based methods, ISTA-Net was trained with a learning rate of 0.0001 and a batch size of 8, whereas MD-DAN used a batch size of 2 with a learning rate of 0.001. MoDL was configured with the number of layers $N$ set to 5 and the number of iterations $K$ to 10. VN employed a network depth of 10 iterations, with a batch size of 10 and a learning rate of 0.001. For diffusion-based methods, DDPM was configured with $\beta_{\text{max}} = 0.02$ and $\beta_{\text{min}} = 0.0001$, and Con-DDPM adopted the same settings. AdaDiff was set with $\beta_{\text{max}} = 20$ and $\beta_{\text{min}} = 0.1$. ScoreMRI was configured with $\sigma_{\text{min}} = 0.01$ and $\sigma_{\text{max}} = 3.78$, using a batch size of 1 and a learning rate of $2 \times 10^{-4}$. PPN employed the same learning rate of $2 \times 10^{-4}$, with the noise range set to $\sigma_{\text{min}} = 0.0001$ and $\sigma_{\text{max}} = 0.02$. For Schrödinger bridge methods, FDB was trained with a batch size of 1 and a learning rate of $1 \times 10^{-4}$. $\mathrm{I}^2$SB-Recon used a batch size of 32 with $\beta_{\text{max}} = 0.3$ and $\beta_{\text{min}} = 1 \times 10^{-5}$, and $\mathrm{I}^2$SB-Inversion adopted the same configuration as $\mathrm{I}^2$SB-Recon. In
The whole process, except the Inversion process, $\eta$ is set
to 1.


\subsubsection{Performance Evaluation}
Three metrics were used to quantitatively evaluate the results, including normalized root mean squared error (NRMSE), peak signal-to-noise ratio (PSNR), and structural similarity index (SSIM)\cite{SSIM}.

\subsection{Experimental Results}


\subsubsection{Guided Reconstruction Experiments}
Fig.~\ref{fig:Guided_knee} shows the results of different guided reconstruction methods on the knee dataset with R = 11.2 and 14.4. At R = 11.2, detail loss can be observed in the reconstructions from JVN, MD-DAN, and Con-DDPM, as indicated by the arrows. At the higher acceleration of R = 14.4, aliasing artifacts appear in the reconstructions of JVN, MD-DAN, and Con-DDPM, with particularly severe artifacts in Con-DDPM. In contrast,  $\mathrm{I}^2$SB-Recon achieves superior performance by effectively removing aliasing artifacts while preserving fine structural details.  $\mathrm{I}^2$SB-Inversion further enhances the performance of  $\mathrm{I}^2$SB-Recon, yielding lower reconstruction errors and higher PSNR at both acceleration rates. Table~\ref{table:guide_knee} summarizes the mean NRMSE, PSNR, and SSIM across 480 knee slices in the guided reconstruction experiments. $\mathrm{I}^2$SB-Inversion achieves the lowest NRMSE and the highest PSNR and SSIM among all compared methods.

\begin{table}[!t] 
\caption{\label{tab: In-Distribution expriments on knee dataset}Guided reconstruction experiments on the knee dataset. The average quantitative metrics were calculated across 480 knee slices at R = 11.2 and R = 14.4.} 
\centering 
\resizebox{\linewidth}{!}{ 
\begin{tabular}{c|cccc} 
\hline \hline 
AF & Method & NRMSE (\%) & PSNR (dB) & SSIM \\ 
\hline 
\multirow{5}{*}{11.2} 
& JVN & 9.40 $\pm$ 2.20 & 34.65 $\pm$ 2.30 & 93.80 $\pm$ 2.85 \\
& MD\text{-}DAN & 8.11 $\pm$ 2.00 & 35.59 $\pm$ 2.10 & 93.83 $\pm$ 2.25 \\
& Con\text{-}DDPM & 9.85 $\pm$ 2.42 & 34.55 $\pm$ 2.36 & 93.75 $\pm$ 2.90 \\
& $\mathrm{I}^2$SB\text{-}Recon & 8.03 $\pm$ 1.95 & 35.78 $\pm$ 2.04 & 93.89 $\pm$ 2.15 \\
& $\mathbf{I^2}\mathbf{SB}\textbf{-Inversion}$ & \textbf{7.51 $\pm$ 1.68} & \textbf{36.41 $\pm$ 1.99} & \textbf{93.95 $\pm$ 1.76} \\
\hline
\multirow{5}{*}{14.4}
& JVN & 11.45 $\pm$ 2.05 & 33.11 $\pm$ 2.38 & 91.94 $\pm$ 2.95 \\
& MD\text{-}DAN & 10.34 $\pm$ 1.85 & 34.12 $\pm$ 2.22 & 92.00 $\pm$ 2.30 \\
& Con\text{-}DDPM & 11.51 $\pm$ 2.22 & 32.78 $\pm$ 2.47 & 91.84 $\pm$ 3.00 \\
& $\mathrm{I}^2$SB\text{-}Recon & 10.26 $\pm$ 1.79 & 34.21 $\pm$ 2.12 & 92.05 $\pm$ 2.20 \\
& $\mathbf{I^2}\mathbf{SB}\textbf{-Inversion}$ & \textbf{9.52 $\pm$ 1.32} & \textbf{34.64 $\pm$ 1.87} & \textbf{92.09 $\pm$ 1.91} \\
\hline \hline 
\end{tabular} 
\label{table:guide_knee} } 
\end{table}

\begin{table}[!t]
   \caption{\label{tab:direct_brain}Direct reconstruction experiments on the brain dataset. The average quantitative metrics were calculated across 1086 brain slices at R = 11.2 and R = 14.4.}
   \centering
   \resizebox{\linewidth}{!}{
       \begin{tabular}{c|cccc}
         \hline \hline AF & Method & NRMSE (\%) & PSNR (dB) & SSIM (\%) \\
         \hline 
         \multirow{10}{*}{11.2} 
& CG-SPIRIT & 10.72 $\pm$ 4.11 & 34.28 $\pm$ 2.83 & 93.65 $\pm$ 3.35 \\
& ISTA-Net & 6.54 $\pm$ 1.94 & 38.52 $\pm$ 1.58 & 94.63 $\pm$ 2.85 \\
& MoDL & 9.06 $\pm$ 3.58 & 36.12 $\pm$ 2.19 & 95.62 $\pm$ 2.35 \\
& VN & 6.68 $\pm$ 2.07 & 38.19 $\pm$ 1.69 & 95.76 $\pm$ 2.22 \\
& PPN & 9.94 $\pm$ 4.06 & 35.93 $\pm$ 2.62 & 94.08 $\pm$ 3.21 \\
& DDPM & 8.12 $\pm$ 2.64 & 37.39 $\pm$ 1.88 & 95.74 $\pm$ 2.18 \\
& Score-MRI & 8.39 $\pm$ 2.76 & 37.03 $\pm$ 1.96 & 95.61 $\pm$ 2.21 \\
& AdaDiff & 7.78 $\pm$ 2.38 & 37.71 $\pm$ 1.79 & 95.39 $\pm$ 2.30 \\
& FDB & 8.93 $\pm$ 3.41 & 36.28 $\pm$ 2.04 & 94.16 $\pm$ 2.74 \\
& $\mathbf{I^2}\mathbf{SB}\textbf{-Inversion}$ & \textbf{6.49 $\pm$ 1.86} & \textbf{38.56 $\pm$ 1.54} & \textbf{96.63 $\pm$ 1.72} \\
\hline
\multirow{10}{*}{14.4}
& CG-SPIRIT & 13.12 $\pm$ 5.81 & 32.61 $\pm$ 2.84 & 91.05 $\pm$ 3.50 \\
& ISTA-Net & 8.91 $\pm$ 2.72 & 36.10 $\pm$ 1.69 & 94.16 $\pm$ 2.40 \\
& MoDL & 12.42 $\pm$ 4.96 & 33.61 $\pm$ 2.48 & 93.17 $\pm$ 2.60 \\
& VN & 9.28 $\pm$ 2.87 & 35.46 $\pm$ 1.81 & 94.03 $\pm$ 2.34 \\
& PPN & 12.90 $\pm$ 5.63 & 33.44 $\pm$ 2.71 & 90.44 $\pm$ 3.65 \\
& DDPM & 10.98 $\pm$ 3.79 & 34.91 $\pm$ 2.02 & 93.32 $\pm$ 2.66 \\
& Score-MRI & 11.32 $\pm$ 3.96 & 34.62 $\pm$ 2.10 & 93.94 $\pm$ 2.42 \\
& AdaDiff & 10.73 $\pm$ 3.58 & 35.06 $\pm$ 1.94 & 93.57 $\pm$ 2.50 \\
& FDB & 11.97 $\pm$ 4.77 & 33.90 $\pm$ 2.29 & 92.36 $\pm$ 2.92 \\
& $\mathbf{I^2}\mathbf{SB}\textbf{-Inversion}$ & \textbf{8.57 $\pm$ 2.39} & \textbf{36.21 $\pm$ 1.61} & \textbf{95.33 $\pm$ 2.05} \\
         
         \hline \hline
       \end{tabular}
   }
\end{table}



\subsubsection{Direct Reconstruction Experiments}
Fig.~\ref{fig:direct_11.2} shows the results of different direct reconstruction methods on the brain dataset at R = 11.2. The conventional CG-SPIRIT method exhibits aliasing artifacts and noise amplification. The unrolling methods, ISTA-Net, MoDL, and VN, effectively suppress noise but introduce image blurring. The diffusion-based methods, including DDPM, Score-MRI, PPN, and AdaDiff, yield larger reconstruction errors compared with $\mathrm{I}^2$SB-Inversion. It is worth noting that the inferior performance of PPN relative to DDPM primarily arises from using fewer denoising steps to accelerate generation, resulting in slightly reduced reconstruction accuracy. FDB produces aliasing artifacts due to the absence of a guidance image. In contrast, $\mathrm{I}^2$SB-Inversion achieves the best quantitative metrics and superior reconstruction quality. Fig.~\ref{fig:direct_14.4} shows the reconstruction results at R = 14.4. At this high acceleration rate, aliasing artifacts become evident in the reconstructions produced by both unrolling-based and diffusion-based methods. In contrast, $\mathrm{I}^2$SB-Inversion maintains high reconstruction quality and achieves the best quantitative metrics. The average quantitative results with R = 11.2 and 14.4 are summarized in Table~\ref{tab:direct_brain}, where $\mathrm{I}^2$SB-Inversion consistently outperforms all other methods.

\section{Discussion}\label{discussion}

In this study, we propose a guided MRI reconstruction framework, termed $\mathrm{I}^2$SB-Inversion, which integrates the Schrödinger Bridge (SB) formulation with an inversion strategy to enhance structural fidelity in cross-modality reconstruction. The proposed method establishes an explicit, pixel-level probabilistic mapping between the guidance and target domains through the SB framework, enabling direct learning of structural correspondences during the generative process. This formulation enables the preservation of detailed anatomical structures from the guidance image while maintaining the physical data consistency required by the reconstruction task.  Moreover, the introduced inversion strategy alleviates cross-modality misalignment by searching for an approximately aligned guidance representation along the SB trajectory based on the initial reconstruction. This process further enhances anatomical consistency and improves reconstruction quality. Consequently, $\mathrm{I}^2$SB-Inversion achieves high acceleration factors of up to 14.4× and outperforms existing SOTA methods.

\subsection{The Effect of Inversion Strategy}
In guided reconstruction, cross-modal misalignment often introduces artifacts and reduces reconstruction accuracy because the reconstructed image must satisfy constraints from both the guidance image and the undersampled data. In $\mathrm{I}^2$SB-Inversion, the structural discrepancies introduced by the guidance image can be reduced through data consistency, yielding a partially corrected image. Then this result is used as a starting point for inversion to infer a more structurally aligned guidance variable, which reinitializes the sampling process. Through this iterative correction, the guidance and target become progressively aligned, thereby mitigating misalignment artifacts and improving reconstruction performance.

\begin{figure}[!t]
    \centerline{\includegraphics[width=0.5\textwidth]{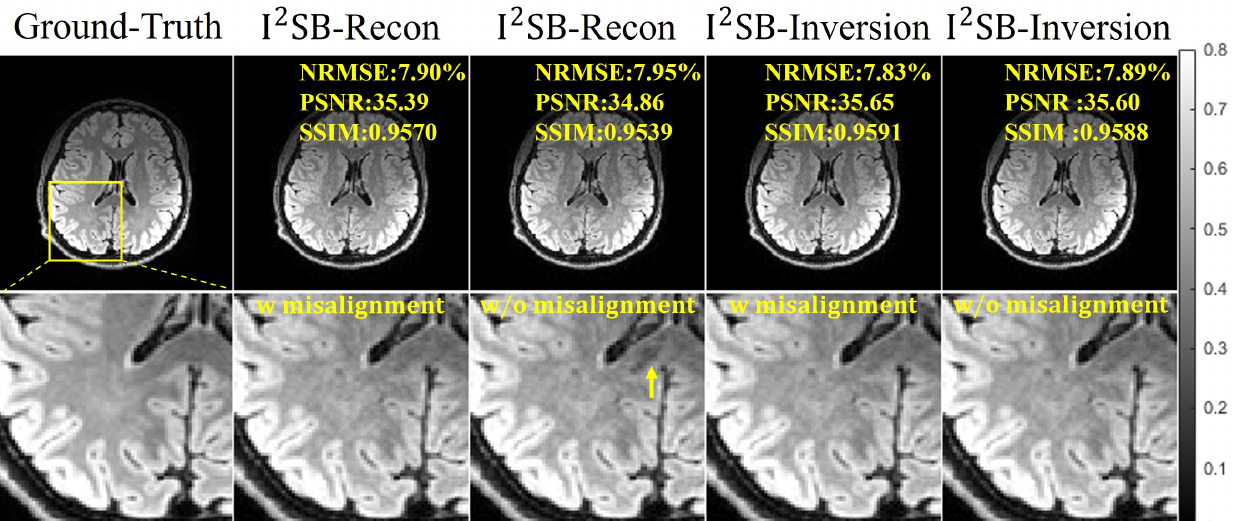}}
    \caption{Reconstruction results under in-plane misalignment at an acceleration factor of R = 14.4. The top row shows the ground-truth and reconstructed images using $\mathrm{I}^2$SB-Recon and $\mathrm{I}^2$SB-Inversion methods w/wo misalignment.
     The bottom row displays enlarged ROIs for detailed comparison, with quantitative metrics reported in yellow and artifacts highlighted by yellow arrows.}
    \label{fig:random_move}
\end{figure}

\begin{figure}[!t]
    \centering
    \includegraphics[width=0.48\textwidth]{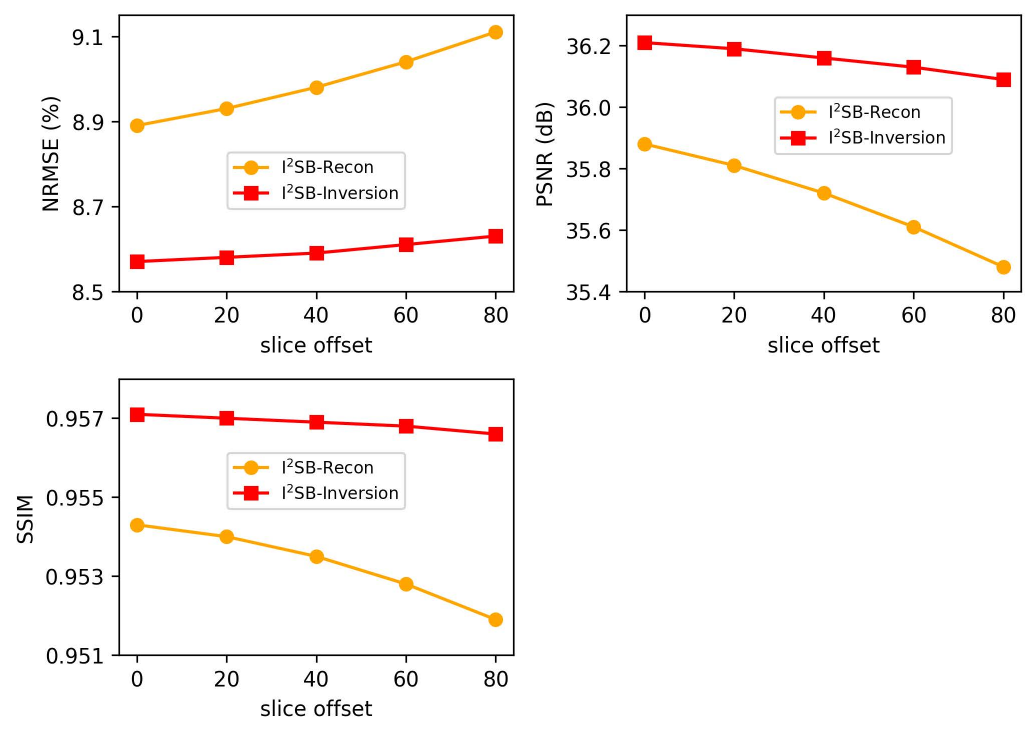}
    \caption{
   Reconstruction performance under structured inter-slice shifts at an acceleration rate of R = 14.4 
    We evaluate NRMSE, PSNR, and SSIM across different levels of simulated slice-wise misalignment. 
    The proposed $\mathrm{I}^2$SB-Inversion method shows high stability with minimal performance degradation as the shift magnitude increases, while baseline $\mathrm{I}^2$SB-Recon is more sensitive to slice offset.}
    \label{fig:slice}
\end{figure}

To assess the effect of the Inversion strategy, two scenarios were simulated with significant misalignment or motion between the guidance and target images. The first scenario simulated in-plane misalignment by translating and rotating the guidance image, where the translations and rotation angles were randomly sampled from Gaussian distributions with zero mean and standard deviations of 10 pixels and 10 degrees, respectively. The second scenario simulated through-plane misalignment by introducing slice offsets between the target and guidance images. For example, when reconstructing the 1st T2 slice, the 1st, 21st, 41st, and 61st T1 slices were used as guidance. 
Fig.~\ref{fig:random_move} presents the reconstruction results of the in-plane misaligned data using $\mathrm{I}^2$SB-Recon and $\mathrm{I}^2$SB-Inversion at R = 14.4. $\mathrm{I}^2$SB-Recon tends to generate pseudo-structural artifacts, whereas $\mathrm{I}^2$SB-Inversion maintains consistent image quality and achieves comparable quantitative performance with or without misalignment. For through-plane misalignment, Fig.~\ref{fig:slice} illustrates the variation of quantitative metrics of the reconstructed images with respect to slice offset. The results show that $\mathrm{I}^2$SB-Inversion exhibits only minor performance degradation as the slice offset increases, whereas the performance degrades significantly without the inversion strategy.

\begin{figure}[!t]
    \centering
    \includegraphics[width=\columnwidth]{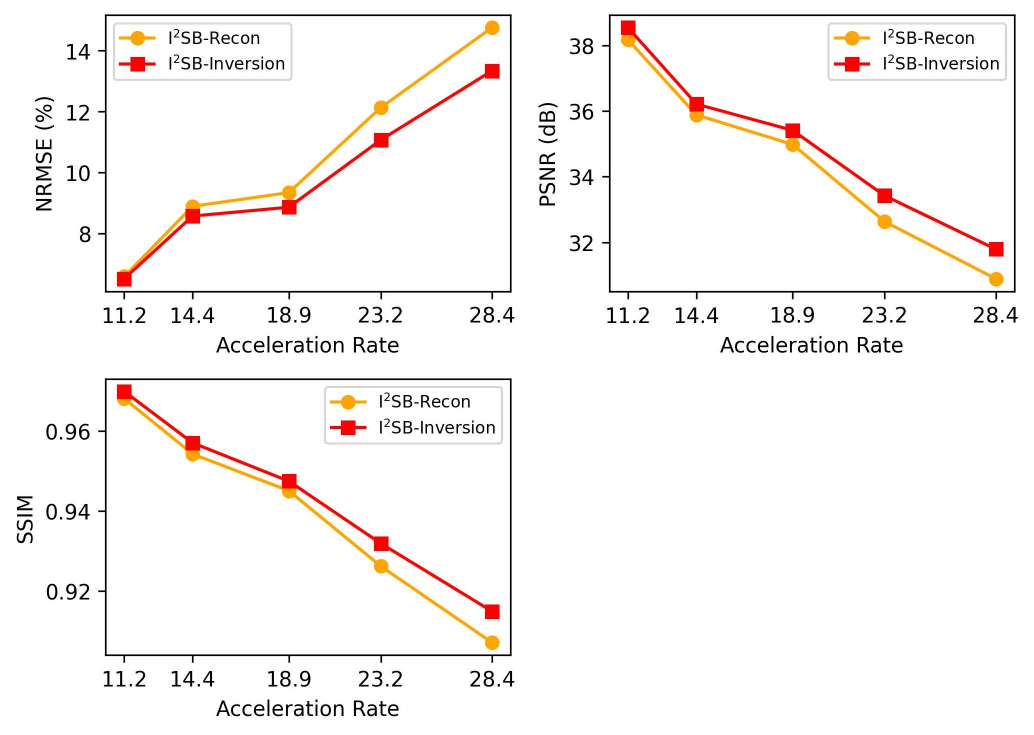}
    \caption{
    The performance comparison between $ \text{I}^2 \text{SB-Recon} $ and $ \text{I}^2 \text{SB-Inversion}$ in terms of NRMSE, PSNR, and SSIM metrics under acceleration rates R = 11.2, 14.4, 18.9, and 23.2.
}
    \label{fig:ablation}
\end{figure}

Furthermore, Fig.~\ref{fig:ablation}  presents the mean NRMSE, PSNR, and SSIM curves of 1086 brain slices reconstructed using  $\mathrm{I}^2$SB-Recon and  $\mathrm{I}^2$SB-Inversion with R = 11.2, 14.4, 18.9, 23.2, and 28.4. It can be observed that the Inversion strategy consistently improves reconstruction performance across all acceleration rates, with the advantage becoming increasingly evident at higher accelerations. This is because subtle structural discrepancies always exist between multimodal images due to inherent differences in imaging physics. At high acceleration rates, the reduced amount of k-space data provides weaker constraints on reconstruction, making the results more vulnerable to these discrepancies. The Inversion strategy can correct these discrepancies and further improve reconstruction accuracy.

\subsection{Extension of Multi-modal Guided Reconstruction}
Eq.~\eqref{MR optimization problem} can be generalized to represent a common multi-modal guided reconstruction problem, where \( \mathbf{y} \), \( \mathbf{x} \), and \( \mathbf{b} \) denote, respectively, the observed signal, the signal to be reconstructed, and the guiding signal. In reconstruction, two main types of challenges are often encountered:

\subsubsection{Problems with Known Forward Operator \( \mathbf{A} \) (e.g., Medical Image Reconstruction)} When the forward operator \( \mathbf{A} \) is known, as in medical image reconstruction tasks, $\mathrm{I}^2$SB-Inversion can be applied across a variety of guided reconstruction models. Typical scenarios include MRI, CT\cite{liu2023dolce}, and ultrasound reconstruction\cite{zhang2023ultrasound}. In such cases, other structural medical images (e.g., PET) can be used as guidance to assist with reconstruction modes that require longer acquisition times, without incurring additional scanning burdens. For example, PET-CT systems can acquire PET and CT data simultaneously,makingg them well-suited for multi-modal imaging needs.

\subsubsection{Problems with Unknown or Non-invertible Forward Operator \( \mathbf{A} \)} In some cases, the forward operator \( \mathbf{A} \) may be unknown or non-invertible, such as in blind image restoration\cite{chung2023parallel}, where reconstruction occurs without a known imaging model. For such cases, approximate solutions can be achieved by appropriately designing inverse problems or by optimizing with a specific loss function via gradient descent.

In summary, I{\( ^2 \)}SB-Inversion extends to diverse applications with known or approximate forward operators, achieving high-precision multi-modal image translation by aligning guiding and target image structures.

\subsection{Reconstruction Time}
Table~\ref{tab:inference_time} summarizes the average reconstruction time of all reconstruction methods. Unrolling-based methods exhibit a clear advantage in inference efficiency. ISTA-Net, MoDL, VN, and JVN all achieve reconstruction times of less than one second per slice, with VN and JVN being the fastest at only 0.06s and 0.08s, respectively. In contrast, diffusion-based models are generally much slower: DDPM and Con-DDPM require approximately 40–50s, while Score-MRI and AdaDiff exceed 100s and 270s, respectively, due to Score-MRI requiring multiple denoising diffusion steps, and AdaDiff using a two-stage diffusion process and prior adaptation, which increases the computational burden. FDB reduces the reconstruction time to 7.12s by constraining the sampling trajectory in the Fourier domain, achieving a more balanced trade-off.
However, its reconstruction performance is limited and cannot match that of more advanced diffusion-based approaches. For the $\mathrm{I}^2$SB framework, superior reconstruction accuracy is achieved at the cost of longer reconstruction time. Specifically, $\mathrm{I}^2$SB-Recon requires 65.21s on average, whereas $\mathrm{I}^2$SB-Inversion is further increased to 122.56s due to the use of an inversion strategy with an additional round of sampling in reconstruction.

\begin{table}[!t]
   \caption{Reconstruction time (in seconds) of different reconstruction methods. Averaged over 100 slices.}
   \centering
   \tiny
   \resizebox{0.8\linewidth}{!}{
       \begin{tabular}{c|c}
         \hline \hline
         Method & Reconstruction  Time (s) \\
         \hline
         CG-SPIRIT & 19.13 \\
         ISTA-Net & 0.30 \\
         MoDL & 0.76 \\
         VN & 0.06 \\
         JVN & 0.08 \\
         MD-DAN & 2.63 \\
         DDPM & 37.98 \\
         Con-DDPM & 49.37 \\
         Score-MRI & 104.28 \\
         AdaDiff & 271.42 \\
         PPN & 6.47 \\
         FDB & 7.12 \\
I$^{\scalebox{0.6}{2}}$SB-Recon & 65.21 \\
I$^{\scalebox{0.6}{2}}$SB-Inversion & 122.56 \\

         \hline \hline
       \end{tabular}
   }
   \label{tab:inference_time}
\end{table}



\subsection{Comparison with Previous Reconstruction Methods Using Cross-Modality Priors}
For reconstruction methods that leverage cross-modality priors, traditional approaches typically construct explicit sparsity or low-rank constraints across multiple contrasts\cite{yang2016sparse,bilgic2011multi,haldar2013improved}. For example, joint-sparsity methods enforce a shared sparse representation in a common transform domain, e.g., group or concatenated wavelets, with coupled $\ell_1$ penalties\cite{bilgic2011multi}, while joint low-rank methods stack multi-contrast images or patches into a Casorati matrix and impose low-rank constraints to capture shared anatomical structures\cite{haldar2013improved}. However, these methods rely on handcrafted priors, which are not trivial and require high computational cost.

Previous deep learning approaches often incorporate cross-contrast priors through learned feature-level fusion\cite{polak2020joint,yang2020model, 10645705}. For example, JVN employs a shared variational reconstruction model that exchanges anatomical information across contrast-specific branches\cite{polak2020joint}. Model-driven attention networks, e.g., MD-DAN and related dual-domain/attention models, explicitly learn the guidance contrast features using attention mechanisms\cite{yang2020model} or dual-domain fusion modules\cite{10645705}. Although learned fusion methods can capture richer, data-driven relationships, they mainly model coarse structural information, and their performance relies heavily on the network architecture and fusion strategy.

In contrast, the proposed SB-based framework offers a pixel-level guidance mechanism that explicitly models the probabilistic translation between the guidance and target modalities. Unlike feature fusion networks that implicitly share information at the feature level, our approach enables direct pixel-wise correspondence and probabilistic coupling between contrasts, leading to more precise structural alignment and preservation of modality-specific details. This pixel-level guidance allows the model to leverage complementary information more effectively while avoiding the over-smoothing issues commonly observed in prior fusion-based methods.

\subsection{Limitation and Future work}

There are several limitations in this study. First, the current implementation focuses on reconstructing a single undersampled target contrast (e.g., T2-weighted) guided by a fully sampled reference contrast (e.g., T1-weighted). Extending this framework to simultaneously reconstruct multiple undersampled contrasts with the guidance image is a natural direction for future work. To achieve this, we plan to design multiple parallel Schrödinger Bridge branches to enable collaborative reconstruction across multiple contrasts. Furthermore, inter-contrast correlations among the reconstructed images can be exploited to further improve reconstruction quality.

Second, the inference process of $\mathrm{I}^2$SB-Inversion involves two stages—initial generation followed by inversion-based refinement—which leads to higher computational cost compared with unrolling-based or other deep learning–based reconstruction methods. Future work will accelerate the sampling process of Schrödinger bridge-based methods to improve their efficiency. Recent advances, such as DPM-Solver++\cite{ DPM-Solver++}, UniP\cite{ UniPC }, Progressive Distillation\cite{ Progressive}, and Flow Matching\cite{ Flow_Matching }, demonstrate that high-fidelity generation can be achieved with only a few or even a single sampling step by optimizing trajectories or using lightweight solvers. Incorporating these fast sampling strategies is expected to improve the practicality of the proposed Schrödinger bridge–based reconstruction.

\section{Conclusions}\label{conclusions}
This study presents $\mathrm{I}^2$SB-Inversion, a Schrödinger Bridge-based framework for guided MRI reconstruction that enables pixel-level cross-modality mapping between different image contrasts. An Inversion strategy is incorporated along the SB path to search for approximately aligned guidance variables based on the initial reconstruction, mitigating the effects of spatial misalignment. Experimental results on paired T1- and T2-weighted datasets demonstrate that $\mathrm{I}^2$SB-Inversion improves reconstruction quality both qualitatively and quantitatively, achieving better artifact suppression and detail preservation compared with SOTA methods.

\section{Acknowledgments}
This study was supported in part by the National Natural Science Foundation of China under grant nos. 62531024, 62322119, 52293425, 62125111, 62331028, 62106252, 12026603, 62206273, 62476268; the National Key R\&D Program of China
 under grant nos. 2022YFA1004203, 2023YFA1011403; the Key Laboratory for Magnetic Resonance and Multimodality Imaging of Guangdong Province under grant no. 2023B1212060052; and the Shenzhen Science and Technology Program under grant nos. RCYX20210609104444089, JCYJ20220818101205012, JCYJ20240813155840052.

\bibliographystyle{IEEEtran}
\bibliography{refs}
\end{document}